\documentclass[10pt, conference, letterpaper]{IEEEtran}
\IEEEoverridecommandlockouts

\usepackage{cite}
\usepackage{amsmath,amssymb,amsfonts}
\usepackage{algorithmic}
\usepackage{graphicx}
\usepackage{textcomp}
\usepackage{xcolor}
\usepackage{mathtools}
\usepackage{multirow}
\usepackage{subcaption}
\usepackage{xcolor}
\usepackage{setspace}
\usepackage{graphicx}
\usepackage{url}
\usepackage{lipsum}

\newcommand*{\affaddr}[1]{#1} 
\newcommand*{\affmark}[1][*]{\textsuperscript{#1}}
\newcommand*{\email}[1]{#1}

\let\OLDthebibliography\thebibliography
\renewcommand\thebibliography[1]{
  \OLDthebibliography{#1}
  \setlength{\parskip}{0pt}
  \setlength{\itemsep}{0.4em}
}

\usepackage{comment}

\begin{document}

\title{Towards Channel-Resilient CSI-Based RF Fingerprinting using Deep Learning}

\author{
Ruiqi Kong\affmark[1] and He (Henry) Chen\affmark[1,~2]\\
\affaddr{\affmark[1]Department of Information Engineering, The Chinese University of Hong Kong, Hong Kong SAR, China}\\
\affaddr{\affmark[2]Shun Hing Institute of Advanced Engineering, The Chinese University of Hong Kong, Hong Kong SAR, China}\\
\email{E-mail: \{kr020, he.chen\}@ie.cuhk.edu.hk}\\
\thanks{This research was supported in part by project \#MMT 79/22 of Shun Hing Institute of Advanced Engineering, The Chinese University of Hong Kong.}
\vspace{-2em}
}

\maketitle

\begin{abstract}


This work introduces \texttt{DeepCRF}, a deep learning framework designed for channel state information-based radio frequency fingerprinting (CSI-RFF). The considered CSI-RFF is built on \texttt{micro-CSI}, a recently discovered radio-frequency (RF) fingerprint that manifests as micro-signals appearing on the channel state information (CSI) curves of commercial WiFi devices. Micro-CSI facilitates CSI-RFF which is more streamlined and easily implementable compared to existing schemes that rely on raw I/Q samples. The primary challenge resides in the precise extraction of micro-CSI from the inherently fluctuating CSI measurements, a process critical for reliable RFF. The construction of a framework that is resilient to channel variability is essential for the practical deployment of CSI-RFF techniques. \texttt{DeepCRF} addresses this challenge with a thoughtfully trained convolutional neural network (CNN). This network's performance is significantly enhanced by employing effective and strategic data augmentation techniques, which bolster its ability to generalize to novel, unseen channel conditions. Furthermore, \texttt{DeepCRF} incorporates supervised contrastive learning to enhance its robustness against noises. Our evaluations demonstrate that \texttt{DeepCRF} significantly enhances the accuracy of device identification across previously unencountered channels. It outperforms both the conventional model-based methods and standard CNN that lack our specialized training and enhancement strategies.

\end{abstract}
%
%
\section{Introduction}

The rapid expansion of low-cost wireless devices under the Internet of Things (IoT) paradigm has introduced substantial security 
challenges for wireless networks \cite{jagannath2022comprehensive}. 
In this context, the development of a unique identifier, or fingerprint, predicated upon the physical layer characteristics of these devices, becomes increasingly vital. RF fingerprinting has been regarded as a promising supplement to existing cryptographic-based security mechanisms for device identification and authentication \cite{angueiraSurveyPhysicalLayer2022}. RF fingerprinting exploits the unique, inherent imperfections in the RF circuitry of wireless transmitters. These hardware imperfections, despite their subtlety, are remarkably consistent and difficult to mimic or tamper with \cite{zhang}. 

RF fingerprinting has demonstrated its capability to uniquely identify devices by using hardware imperfections arise from manufacturing processes, even when devices are transmitting identical messages. Such imperfections lead to signal distortions that, while typically having a negligible effect on the decoding of transmitted data, provide a sufficiently distinctive fingerprint for each device. This allows for robust identification, distinguishing between devices produced by the same manufacturer due to the inherent variability in hardware production. Furthermore, RF fingerprinting offers a significant advantage in terms of computational efficiency. Unlike cryptographic methods that demand considerable resources for key generation and management, RF fingerprinting leverages the intrinsic properties of the RF hardware for identification \cite{9450821}. This approach not only ensures a robust security solution but also saves resources in the IoT ecosystem.

Previous studies have established that it is viable to identify wireless devices based on characteristic imperfections in their RF circuitry, such as sampling frequency offset, transient signals, I/Q imbalance, nonlinearity of power amplifier of wireless transmitters \cite{Brik, transient, dac, Remley2005,Sheng2008}. Given the challenges in explicitly and accurately modeling the subtle nonlinear effects intrinsic to RF hardware, deep learning emerges as an intuitive solution for extracting device fingerprints. Subsequent advancements have demonstrated that convolutional neural networks (CNN) can significantly enhance the accuracy of such device fingerprinting, outperforming traditional model-based methods \cite{merchant2018deep, cekic2021wireless, shen2023towards}. The introduction of CNN has boosted the ability to accurately identify devices with the presence of environmental changes. However, the abovementioned fingerprinting methods require access to physical-layer signal samples and additional instruments for I/Q sample collection. This requirement largely limits their practical application in commercial off-the-shelf (COTS) systems.

Thanks to developments in WiFi channel state information (CSI) tools\cite{atheros,linux,picoscenes}, COTS WiFi devices can report the CSI measurements captured at physical layer to upper layers, making CSI-based RF fingerprinting (CSI-RFF) appealing \cite{hua,liu,lin2020,kong2023physicallayer,meneghello2022deepcsi}. CSI-RFF is more streamlined and easily implemented than radio sample-based solutions. 
Studies \cite{hua,liu,lin2020,kong2023physicallayer} have adopted model-based methods to extract various fingerprints from CSI, achieving satisfactory identification accuracy. However, these algorithms typically demand a sizable number of CSI measurements to mitigate noises and channel distortions \cite{liu,kong2023physicallayer,lin2020,hua}. They are either trained and tested in the same environment\cite{hua,liu,lin2020} or tested only in unseen environments under line-of-sight (LoS) conditions\cite{kong2023physicallayer}. Another method\cite{meneghello2022deepcsi} uses deep learning to extract RF fingerprints from compressed beamforming feedback, but it also underperforms in varying channel conditions. Therefore, fluctuating wireless channels and noises pose significant challenges for CSI-RFF.

 \begin{figure}
    \centering
\includegraphics[width=\linewidth]{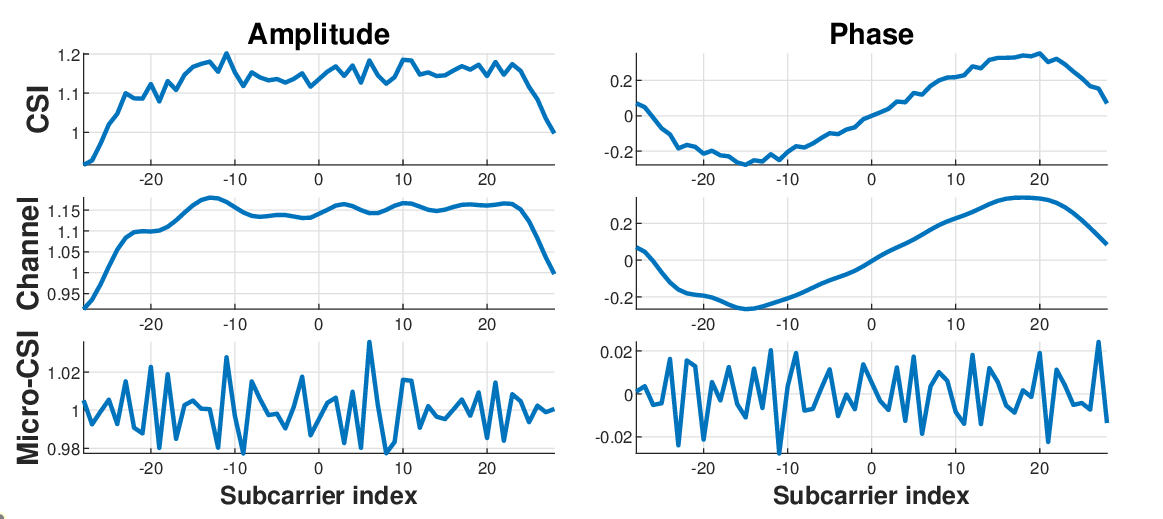}
  \caption{RF fingerprint extracted from denoised LoS CSI  \cite{kong2023physicallayer}.}
  \label{fingerprint}
     \vspace{-1.5em}
\end{figure}

In an effort to address these challenges, we build upon our previous work \cite{kong2023physicallayer} and introduce \texttt{DeepCRF}, a deep learning-enabled framework for micro-CSI-based RF fingerprinting that enhances accuracy in challenging, unseen channel conditions and improves robustness against noises. 
Our proposed \texttt{DeepCRF} comprises an encoder and a classifier, each serving a unique function. A primary challenge lies in ensuring the encoder learns small-scale micro-CSI (i.e., the fingerprint), not large-scale channel information, given the variability of wireless channels. To tackle this, we introduce an effective data augmentation technique tailored for micro-CSI, enabling the network to learn it while reducing extensive data collection requirements. Furthermore, to counter the noise impact on the small-scale micro-CSI, we employ supervised contrastive learning to enhance the noise robustness of the extracted fingerprints. Evaluation on a synthetic testing dataset reveals that at a 40 dB signal-to-noise ratio (SNR), \texttt{DeepCRF} achieves over 97\% identification accuracy among 19 COTS WiFi 4/5/6 devices merely using one CSI sample. In contrast, the model-based approach in \cite{kong2023physicallayer} and a straightforward application of CNN experience a 72\% and 55\% reduction in accuracy, respectively. These results underscore the effectiveness of \texttt{DeepCRF} in extracting reliable and robust RF fingerprints and its potential applicability in real-world scenarios.

\section{Micro-CSI-based RF Fingerprinting}

This section presents the signal model of micro-CSI \cite{kong2023physicallayer}. We also justify, through preliminary simulations, the necessity of adopting deep learning in achieving channel-resilient micro-CSI-based RFF. 

\subsection{Signal Model of Micro-CSI}

The estimated CSI at the receiver side incorporates the distortions induced by both wireless channels and the transmitter’s RF circuitry imperfections\footnote{Our experimental results in \cite{kong2023physicallayer} show that the receiver's RF circuitry imperfections have negligible impacts on micro-CSI.} that can be used as RF fingerprinting. CSI estimation in WiFi protocols is achieved by utilizing the Long Training Symbol (LTS), which is part of the preamble in each packet. Since the LTS is predefined by protocols, hardware-induced distortions, whether linear or nonlinear, will uniformly alter the LTS samples across a specific device. Theoretically, we can view the hardware distortions as deviations applied to the standard LTS samples. Given this model, the estimated CSI output generated by WiFi CSI tools can be expressed as follows \cite{kong2023physicallayer}
\begin{equation}\label{est_CSI}
\Tilde{\bold{c}}= \Tilde{\bold{h}} \circ (\bold{1}+\Tilde{\bold{f}})+\Tilde{\bold{z}} ,
\end{equation}
where $\Tilde{\bold{f}}$ denotes the hardware imperfection-induced fingerprint, $\Tilde{\bold{h}}$ is the discrete-time equivalent channel that captures the joint effects of multipath and filtering, $\circ$ is element-wise multiplication and $\bold{1}$ represents the vector with all 1's, and $\Tilde{\bold{z}}$ is the frequency-domain noise vector. As demonstrated by Eq. (\ref{est_CSI}), the channel information $\Tilde{\bold{h}}$ and the hardware fingerprint $\Tilde{\bold{f}}$ are entangled. Theoretically, due to the inherent sparsity of channels, the number of time-domain taps in wireless channels is often significantly less than the length of an OFDM symbol. Consequently, the fingerprint residing in other unoccupied dimensions can be separated from the channel information. Our prior work\cite{kong2023physicallayer} employs the least squares (LS) estimation method to extract the fingerprint from denoised CSI under strong line-of-sight (LoS) conditions, as depicted in Fig.~\ref{fingerprint}. The denoising process involves averaging multiple CSI measurements gathered in a static environment over a short period.

\subsection{{Why Deep Learning?}}
\label{why}

\begin{figure}
  \centering
    \includegraphics[width=1\linewidth]{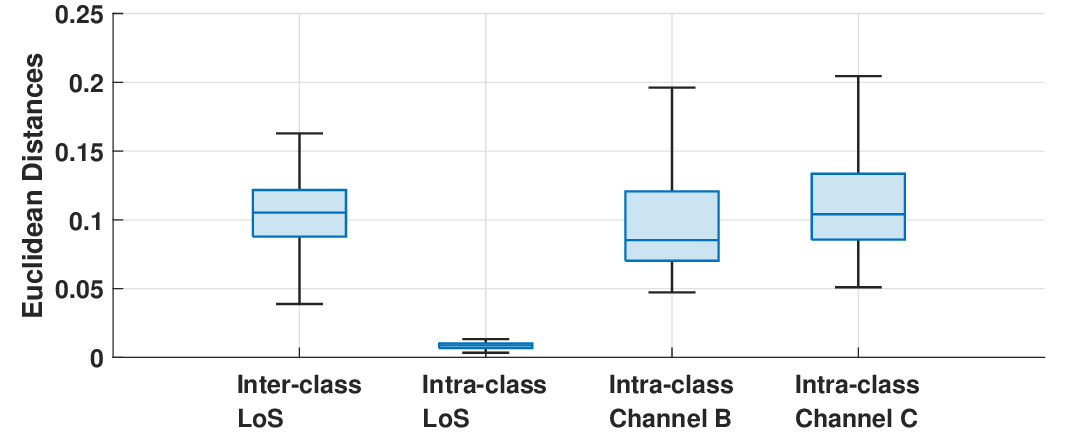}
    \caption{Distances between fingerprints of different NICs and fingerprint extraction errors caused by environmental changes. }
  \label{impact}
  \vspace{-1.5em}
\end{figure}

In this subsection, we conduct a preliminary study to explore the potential limitations of the model-based LS method \cite{kong2023physicallayer} using synthetic CSI data with more complex multipath conditions. Our results showed that the LS method fails to extract fingerprints with high accuracy under complex multipath channels. This limitation motivates us to develop a deep learning-based approach. 

To enhance the authenticity of the synthetic CSI data, we produce them by multiplying real-world CSI measurements of 19 WiFi network interface cards (NICs), collected under LoS conditions, to synthetically generated multipath channels. Detailed descriptions of the CSI data collection and synthesis processes are provided in Sections \ref{sec: eva}-A and \ref{method}-C, respectively. For clarity, henceforth we will distinguish real-world CSI measurements from their synthetic counterparts by referring to the latter as CSI samples. In this preliminary investigation, we employ the LS method to extract fingerprints from both real-world CSI measurements and synthetic CSI samples. Subsequently, we calculate the Euclidean distances of the extracted fingerprints. Fig.~\ref{impact} presents our findings using box plots to illustrate the inter-class and intra-class distances among the fingerprints. The inter-class distances refer to the distances between fingerprints from different NICs, highlighting the identifiability between NICs. On the other hand, intra-class distances, which are derived from CSI pertaining to a single NIC  subject to various channel conditions, characterize the stability of fingerprints despite channel condition changes. 


We first compare the distance variations within inter-class and intra-class under LoS conditions. We use real-world denoised CSI measurements collected in LoS conditions as a baseline against the complex multipath and noisy scenarios. The first two box plots in Fig.~\ref{impact} show that intra-class variations is less than the minimum distinguishable distance between 19 distinct WiFi NICs in LoS conditions. This result supports the effectiveness of the LS method in LoS environments. Note that the observed variability in intra-class distances under LoS conditions can be primarily ascribed to residual noise, which cannot be completely eliminated during the denoising process.


Further into the analysis, we explore the impact of channel conditions on intra-class distances by using synthetic data involving complex multipath scenarios. These scenarios encompass Model-B, replicating an indoor residential space, and Model-C, representing an indoor office environment, both evaluated at a signal-to-noise ratio (SNR) of 40 dB. All synthetic scenarios were simulated using the MATLAB WLAN toolbox \cite{wlantool}. Fig.~\ref{impact} illustrates a marked contrast between the stability of fingerprints extracted under LoS conditions and those obtained in the presence of multipath and noisy channel variations. The latter exhibit fluctuations that exceed the threshold for reliable differentiation between NICs. This suggests that the variability induced by the channel conditions could result in the misidentification of a single NIC as multiple distinct entities. This observation underscores the adverse impact that multipath channels have on the accuracy of CSI-RFF when employing the LS method. An intuitive explanation of this observation is that the presence of more multipath makes it harder for the LS method to separate the channel and the fingerprint. Further, the small scale of micro-CSI makes it  particularly susceptible to noise contamination.
These challenges motivate our design of \texttt{DeepCRF}, a model-inspired deep learning approach for more channel-resilient CSI-RFF.

\section{Method}
\label{method}

This section first overviews the network structure of \texttt{DeepCRF} before elaborating on the design of each component and training strategies.
\subsection{Overview of \texttt{DeepCRF} }

\texttt{DeepCRF} consists of an encoder and a classifier. Each has a unique role, and we explain them using a two-stage training process, shown in Fig.~\ref{architecture}. The primary objective of \texttt{DeepCRF}'s encoder is to extract RF fingerprints resilient to unseen channel conditions. Considering the variability of wireless channels across diverse environments, it becomes imperative to ensure that the encoder learns intrinsic RF distortions within the CSI, as opposed to distortions induced by the wireless channel. To facilitate this, we propose a uniquely tailored data augmentation strategy for micro-CSI, thereby introducing additional variations and diversity to the training dataset, subsequently enhancing the generalization capability of \texttt{DeepCRF}. 

During the initial stage of training (Stage 1), we incorporate a supervised contrastive loss function \cite{khosla2020supervised} for better noise resistance. This loss function prompts the encoder to generate representations that are closer for inputs of the same class while maintaining further distance between representations from different classes. The inclusion of a project head, corroborated by previous research to be beneficial \cite{chen2020simple}, provides auxiliary guidance throughout the training process. In the second stage (Stage 2), our goal is to train a classifier to realize robust device identification. The pre-trained parameters of the encoder's are transferred, providing a foundational understanding for further refinement. The focus then shifts to train the classifier, which is built upon the pre-initialized encoder. Concurrently, the encoder is fine-tuned to synergize with the classifier. Throughout Stage 2, a cross-entropy loss function is utilized to guide the training and fine-tuning process. This two-stage approach ensures that \texttt{DeepCRF} not only learns representative features against noises but also achieves a high accuracy in device identification.

\subsection{Data Augmentation}
\label{da}
\begin{figure}
    \centering
\includegraphics[width=\linewidth]{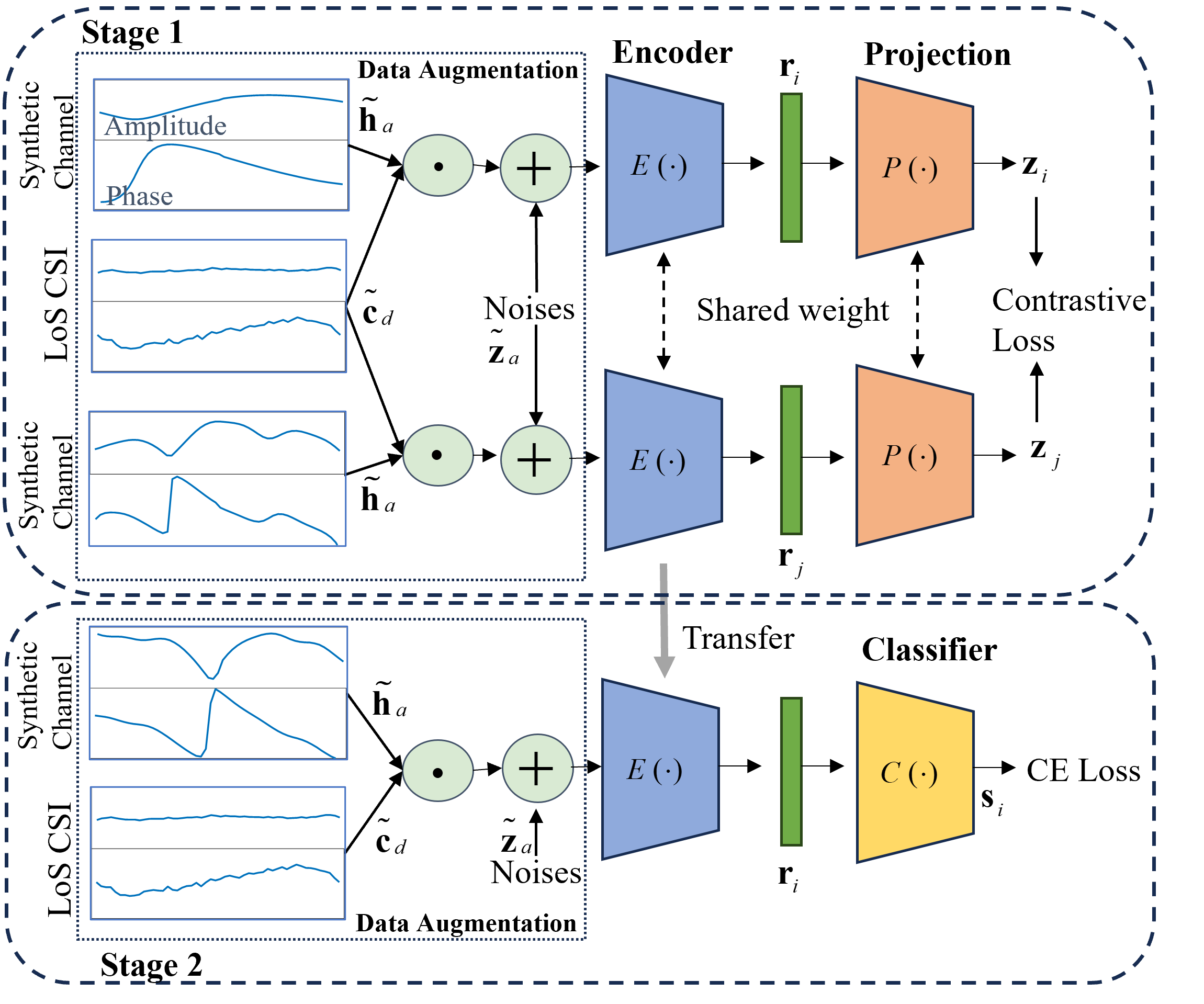}
    \caption{Overview of \texttt{DeepCRF}.}
    \label{architecture}
\vspace{-2em}
\end{figure}

In order for the encoder to effectively learn small-scale RF distortions, it is  advantageous to expose it to a sizeable dataset spanning diverse channel conditions. However, gathering a large-scale practical CSI dataset can be a painstaking process. As a solution, we propose an augmentation strategy rooted in the signal model of micro-CSI. {This approach allows us to synthesize augmented CSI samples representing diverse environments by using real-world CSI measurements collected from a simple, low-effort setup.}

Let $\Tilde{\bold{c}}_d$ denote the denoised CSI, captured by hardware under LoS conditions. Assuming that the noises in the denoised CSI is negligible, we can express the denoised CSI as $\Tilde{\bold{c}}_d \approx \Tilde{\bold{h}} \circ (\bold{1}+\Tilde{\bold{f}})$. In this expression, it is the hardware-induced distortions, symbolized by $\Tilde{\bold{f}}$, from which we desire the encoder to extract features. In contrast, $\Tilde{\bold{h}}$ is subject to fluctuation with the environmental changes, thus lacking the consistency required to act as a reliable identifier across different devices.

Therefore, the augmentation strategy we propose is one that artificially introduces a breadth of channel conditions while maintaining the integrity of $\Tilde{\bold{f}}$. Such a strategy enhances the training dataset, reinforcing the capacity of the encoder to learn device-specific RF fingerprints without being misled by the channel-induced variations. Moreover, as elucidated in Section~\ref{why}, the small-scale nature of micro-CSI renders it particularly sensitive to noises, necessitating the incorporation of noises in the augmentation process to mimic real-world scenarios. The data augmentation process is mathematically formalized as
\begin{equation}\label{aug}
\Tilde{\bold{c}}_a= \Tilde{\bold{h}}_a \circ \Tilde{\bold{c}}_d + \Tilde{\bold{z}}_a\approx (\Tilde{\bold{h}}_a \circ \Tilde{\bold{h}}) \circ (\bold{1}+\Tilde{\bold{f}}) + \Tilde{\bold{z}}_a,
\end{equation}
where $\Tilde{\bold{h}}_a$ denotes the artificially generated channel, and $\Tilde{\bold{z}}_a$ embodies the added synthetic noises. The equation presented in (\ref{aug}) confirms that the synthesized data $\Tilde{\bold{c}}_a$ align with the structure of the signal model defined in (\ref{est_CSI}). This compatibility validates the utility of the synthetic data in augmenting real-world datasets. Furthermore, the channel characteristics embedded within the synthetic data $\Tilde{\bold{c}}_a$ is depicted by the convolution of $\Tilde{\bold{h}}_a$ with $\Tilde{\bold{h}}$ in the time domain, obtaining more severe multipath effect to enrich the dataset. Before inputting into the encoder, each augmented complex-valued CSI sample is decomposed into two real-valued vectors: one representing the amplitude, and the other encapsulating the phase information, as shown in Fig.~\ref{architecture}. 

\subsection{{Neural Network Design}}

We now delve into the designs of the three main components of \texttt{DeepCRF}, namely encoder, projection head, and classifier.

The encoder function $E(\cdot)$ translates the augmented CSI sample $\Tilde{\bold{c}}_a$ into the latent representation $\bold{r}$, i.e., $\bold{r}= E(\Tilde{\bold{c}}_a)$. To achieve this, we propose a variant of ResNet-18 as our encoder, given the original ResNet \cite{resnet} was primarily designed for learning visual data representations. In our variant, we eliminate the final fully connected layer and adjust the initial convolutional layer. This adjusted layer incorporates 64 filters, each sized $1\times3$, which separately capture spatial information in the frequency domain for either amplitude or phase dimension. Considering the interdependency between amplitude and phase in CSI, we add another convolutional layer following the first. This additional layer uses 64 filters of size $2\times3$, allowing it to learn from both amplitude and phase dimensions simultaneously. Furthermore, we substitute the activation function with the Gaussian Error Linear Unit (GELU) to enhance performance.

The projection head $P(\cdot)$ within Stage 1 translates representations $\bold{r}$ into a hypersphere where the contrastive loss is applied for better alignment and uniformity of representations \cite{wang2020understanding}. We use a fully connected layer featuring 128 neurons to obtain the normalized output as $\bold{z} = \frac{P(\bold{r})}{||P(\bold{r})||_2}$ \cite{chen2020simple}. During the second stage, the classifier $C(\cdot)$—which is constructed as a single fully connected layer—assigns the representations $\bold{r}$ to their corresponding class probabilities $\bold{s}$, denoted by $\bold{s}= C(\bold{r})$. The number of neurons in this layer corresponds to the total number of classes.

\subsection{Loss Function}

In Stage 1, we use supervised contrastive loss \cite{khosla2020supervised} to guide the training process to enhance the discriminative power of the learned representations. The contrastive loss $\mathcal{L}^{\text {c}}$ is computed as the sum of individual contrastive losses $\mathcal{L}_{i}^{\text{c}}$ for each arbitrary augmented sample $i$ in the batch $I$. The contrastive loss for a specific sample $i$ is defined as: 
\begin{equation}
\label{loss_eq}
\mathcal{L}_{i}^{\text{c}} = \frac{-1}{|J(i)|} \sum_{j \in J(i)} \log \frac{\exp(\bold{z}_i^T \bold{z}_j / \tau)}{\sum_{k \in K(i)} \exp(\bold{z}_i^T \bold{z}_k / \tau)},
\end{equation}
where $K(i) \equiv I \backslash  \{i\}$, the term $J(i)$ represents the set of indices of all samples with the same label as $i$ in the batch, excluding $i$, and $|J(i)|$ is the cardinality of $J(i)$. 
In Stage 2, the model is guided by cross-entropy (CE) loss, which is well-established for classification tasks.

\section{Evaluation}\label{sec: eva}
In this section, we present a performance assessment of \texttt{DeepCRF}. We begin by detailing the CSI collection setup and the realization of our augmentation strategy. Subsequently, we delve into the specifics of the training implementation. We conclude by analyzing the performance of \texttt{DeepCRF}.

\subsection{Data Collection and Augmentation}

For the evaluation of performance, our experimental setup included 19 WiFi 4/5/6 NICs from three different manufacturers. The inventory comprised 6 MediaTek MT7601, a single MT7612, 7 Atheros AR9271, one Ralink RT3070, an RT8811, an RT8821, an RT8822, and one Intel AX200 NIC. We set up a suite of software-defined radio (SDR) devices to capture CSI from acknowledgment (ACK) packets transmitted by the NICs in LoS configurations. The used SDR consists of a Xilinx Zynq ZC706 development \cite{zc706} and an FMCOMMS5 ADI daughter board \cite{coms5}. The acquired samples were decoded, and their corresponding CSI reports were extracted using a MATLAB 802.11 analysis program \cite{analysis}. The wireless network was set to operate on the 2.4 GHz spectrum, specifically Channel 10, with a 20 MHz bandwidth. The ACK packets transmitted with 52 active subcarriers. The resulting practical dataset comprised a total of 80,833 CSI measurements.

To expand the diversity of the practical dataset, which originally encompassed CSI only from a single position, we augmented the denoised CSI measurements by simulating 6 different types of indoor channel conditions. These conditions included line-of-sight (LoS) and non-line-of-sight (NLoS) scenarios for three indoor channel models: Model-B, Model-C, and Model-D, utilizing the \texttt{wlanTGnChannel} function in accordance with the WLAN toolbox specifications \cite{wlan}. The resultant synthetic dataset contained 912,000 CSI samples. These CSI samples embody different realizations of the 6 specified channel types, which are further diversified by the inclusion of noises across 8 distinct SNR levels, from 5 dB to 40 dB, applying the \texttt{awgn} function. For our experimental evaluations, we segmented the synthetic dataset into three distinct subsets: 80\% for training, 10\% for validation, and the remaining 10\% for testing.

\subsection{Training Details}

We train the network weights using an Adam optimizer, with an initial learning rate of $10^{-4}$ and a weight decay of $10^{-5}$. The batch size is set to 512, and a patience of 10 is used to tolerate non-converging validation loss. For the loss function \eqref{loss_eq}, the temperature parameter $\tau$ is set to 0.07 \cite{khosla2020supervised}. The proposed architecture is implemented in PyTorch and executed on an Ubuntu platform using an NVIDIA RTX A6000 GPU equipped with 48 GB of VRAM. 

\subsection{Identification Performance} 

\begin{figure}
  \centering
  \begin{subfigure}{\linewidth}
   \includegraphics[width=\linewidth]{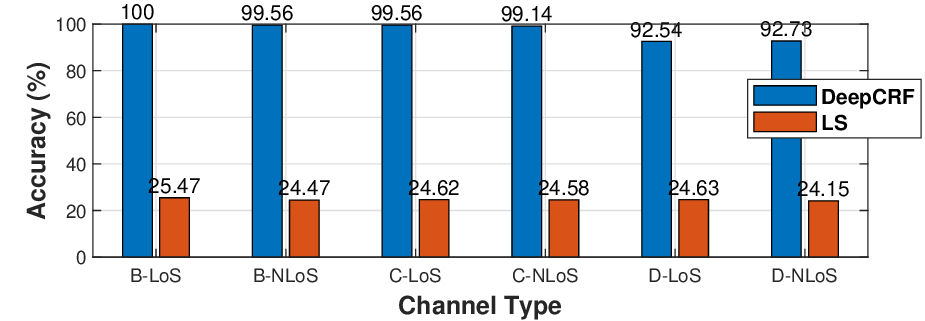}
   \caption{Multipath impact, when SNR $=40 dB$.} 
   \end{subfigure}
   \begin{subfigure}{\linewidth}
   \includegraphics[width=\linewidth]{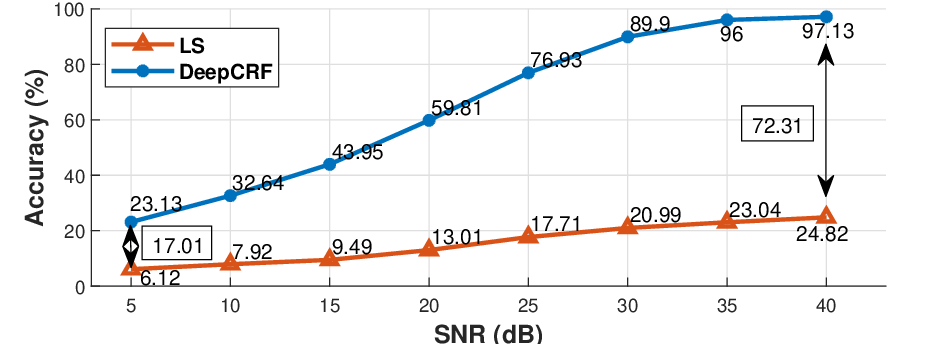}
   \caption{SNR impact, when considering all 6 channel types.} 
   \end{subfigure}
  \caption{Identification accuracy (\%) of model-based method (LS \cite{kong2023physicallayer}) and deep learning-based method (\texttt{DeepCRF}).}
  \label{classification}
  \vspace{-2em}
\end{figure}

In this analysis, we compared the performance of the \texttt{DeepCRF} and the model-based LS method \cite{kong2023physicallayer} for device identification using a single CSI sample under a range of challenging, previously unencountered channel conditions at various SNR levels. To ensure a fair comparison, both methods incorporated a fully connected layer for identification purposes. The LS method, which was trained on the real-world dataset obtained under LoS conditions and evaluated against a synthetic dataset with multipath configurations. As illustrated in Fig.~\ref{classification}a, the LS method underperformed in all channel conditions compared to \texttt{DeepCRF}. This significant deficit highlights the LS method's vulnerability to multipath interference, impairing its channel fingerprinting precision. Conversely, \texttt{DeepCRF} demonstrated a distinct superiority in identical test scenarios, maintaining an accuracy exceeding $92\%$ even in complex indoor NLoS settings. Furthermore, Fig.~\ref{classification}b reveals \texttt{DeepCRF}'s heightened noise immunity. Notably, \texttt{DeepCRF} maintains superior performance over the LS method at all SNR levels tested, with the performance gap expanding as SNR increases.
These results suggest that \texttt{DeepCRF} maintains robust and reliable device identification performance, even under challenging and unseen channel conditions and showcases a remarkable resistance to noises.

\begin{table}[]
  \caption{Results of ablation studies on strategies: Supervised Contrastive Learning (SCL) and Data Augmentation (DA)}
  \label{performance}
  \label{a2}
  \centering
  \renewcommand\arraystretch{1.5}
\resizebox{\linewidth}{!}{
  \begin{tabular}{l|cccccccc}
    \hline
    Method & 5dB&10dB&15dB&20dB&25dB&30dB&35dB&40dB \\
    \hline  
  \texttt{DeepCRF}          & \textbf{23.13} & \textbf{32.64} & \textbf{43.95}& \textbf{59.81} & \textbf{76.93}&\textbf{89.90}&\textbf{96.00}&\textbf{97.13}\\
  \hline
   w/o SCL   & 22.76 &32.06& 42.43& 57.09 &74.59 &87.48& 93.78& 95.33 \\
   w/o DA  & 6.18& 6.81 &8.05& 11.58 &20.59 &32.53& 40.89 &44.65 \\
   w/o (DA+SCL)  & 5.36& 6.08 &7.55& 10.38 &16.98 &28.99& 38.71 &42.62 \\
  \hline
\end{tabular}
}
  \vspace{-1em}
\end{table}

\begin{figure}
  \centering
  \begin{subfigure}{.49\linewidth}
      \includegraphics[width=\linewidth]{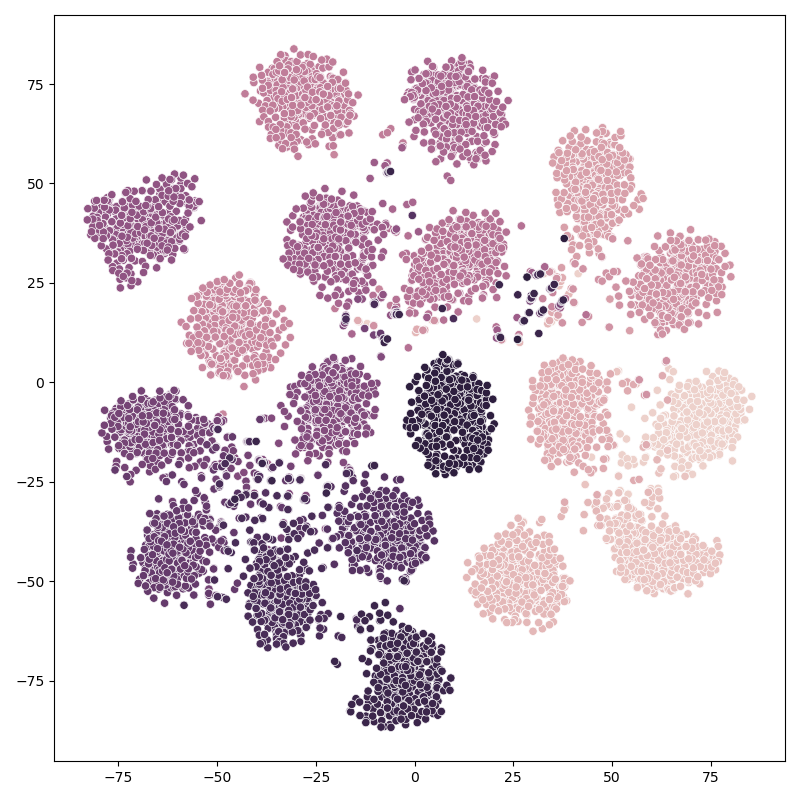}
      \caption{Standard Cross-Entropy Loss}
  \end{subfigure}
    \begin{subfigure}{.49\linewidth}
    \includegraphics[width=\linewidth]{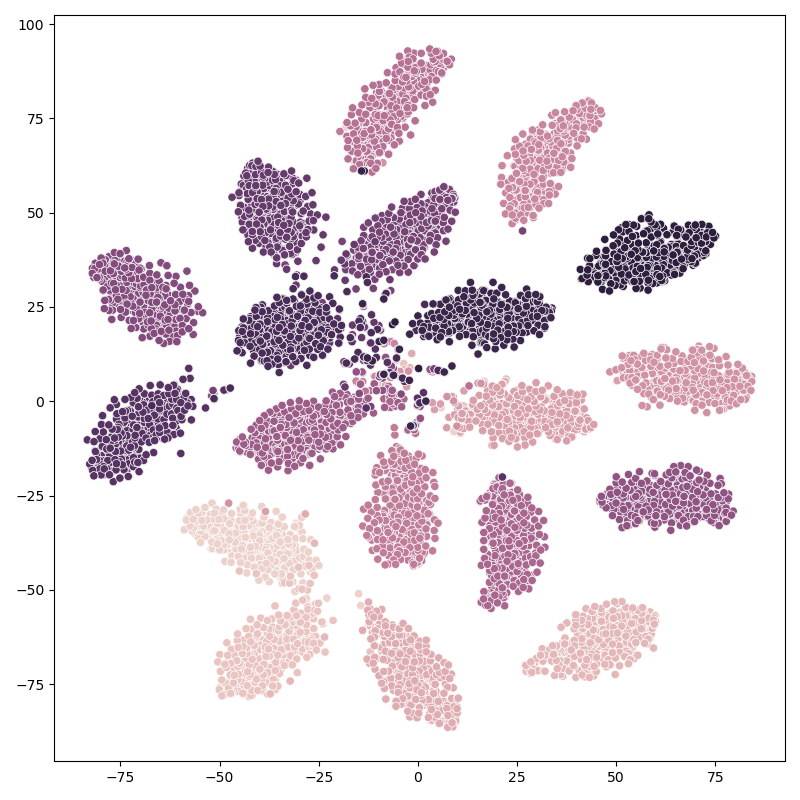}
      \caption{Supervised Contrastive Loss}      
  \end{subfigure}
  \caption{Visualization of the learned representations.}
  \label{tsne}
    \vspace{-1.5em}
\end{figure}
\vspace{-0.5em}
\subsection{Ablation Studies}
The training strategies for \texttt{DeepCRF} have proven to be pivotal in obtaining superior accuracy, as evidenced by the following ablation studies. As Table~\ref{performance} illustrates, incorporating supervised contrastive learning (SCL) into the training regimen yields, an average of 1.75\% boost in accuracy across all SNR levels. To graphically represent the high-dimensional data learned by \texttt{DeepCRF}, we employed t-Distributed Stochastic Neighbor Embedding (t-SNE) \cite{van2008visualizing}. This technique allows us to project the 512-dimensional encoder outputs into a 2D plane, with each data point colored according to its true label, as depicted in Fig.~\ref{tsne}. This visualization demonstrates that data points corresponding to the same label cluster more tightly when trained with contrastive loss, while those with different labels are more distinctly separated, in comparison to training with cross-entropy loss. Such clustering indicates that the supervised contrastive loss used in the initial training stage enhances class separability, which in turn contributes to the improved noise resistance of the model.

Besides, the performance of \texttt{DeepCRF} suffers significantly without the implementation of data augmentation (DA), achieving a maximum accuracy of just 44.65\% at 40 dB SNR. This emphasizes the essential role of exposing the model to CSI data from diverse channel conditions during training. This exposure allows \texttt{DeepCRF} to effectively discern and capture the RF fingerprints from the CSI data, rather than the channel distortions. Furthermore, this underscores the effectiveness of our data augmentation strategy in bolstering the training process. Additionally, when \texttt{DeepCRF} is trained without the benefits of both DA and SCL, there is a further decline in performance, underscoring the combined contribution of both strategies to the robustness and accuracy of the model.
\section{Conclusions}
In this paper, we present \texttt{DeepCRF}, a new deep learning-enabled framework tailored for CSI-based radio-frequency fingerprinting. \texttt{DeepCRF} was evaluated on a synthetic dataset encompassing diverse indoor environments. The synthetic dataset is constructed by augmenting real-world line-of-sight CSI measurements with simulated multipath effects and noises that generated using the MATLAB toolbox. This evaluation demonstrated \texttt{DeepCRF}'s remarkable resistance to channel variations and noises. Specifically, at a signal-to-noise ratio of 40 dB and merely using a single CSI sample, \texttt{DeepCRF} surpassed the device identification accuracy of an existing model-based method by 72\%. The significant enhancement in the performance of \texttt{DeepCRF} is largely due to our proposed data augmentation approach, which thoughtfully incorporates the specific attributes of the adopted RF fingerprint, coupled with the integration of supervised contrastive learning for enhanced noise resistance. Our approach's effectiveness has been substantiated through experiments with synthetic CSI data. Future work will include further evaluation of \texttt{DeepCRF} on real-world CSI measurements to be collected under complex indoor environments.

\bibliographystyle{ieeetr}
\bibliography{refs}

\end{document}